%% file: main.tex
\documentclass[conference]{IEEEtran}
\IEEEoverridecommandlockouts

\input{preamble}

\begin{document}
\title{A brief overview of programmed instructions for quantum software education}

\author{\IEEEauthorblockN{Richard A. Wolf}
\IEEEauthorblockA{\textit{Department of Computer Science} \\
\textit{University of Galway}\\
\textit{Irish Center for High-End Computing}\\
Galway, Ireland}
\and
\IEEEauthorblockN{Sho Araiba}
\IEEEauthorblockA{\textit{Psychology Department} \\
\textit{University of Hawaii}\\
Honolulu, United States}
}

\maketitle

\input{sections/abstract} 


\input{sections/intro} 
\input{sections/what_is_pi} %
\input{sections/pi_for_qed} %
\input{sections/example_pi} %
\input{sections/conclusion} %

\input{sections/acknowledgments}

{\small
\bibliographystyle{abbrv}
\bibliography{main}
}

\input{sections/annex}

\end{document}

%% file: preamble.tex
\usepackage{cite}
\usepackage{amsmath,amssymb,amsfonts, txfonts}
\usepackage{algorithmic}
\usepackage{graphicx}
\usepackage{textcomp}
\usepackage{xcolor}
\usepackage{braket}
\usepackage{booktabs}
\usepackage{multirow}
\usepackage{pifont}
\usepackage{tablefootnote}
\usepackage{caption, subcaption}
\usepackage[export]{adjustbox}
\usepackage{listing}

\newcommand{\tnl}{T\&L }%

\def\BibTeX{{\rm B\kern-.05em{\sc i\kern-.025em b}\kern-.08em
    T\kern-.1667em\lower.7ex\hbox{E}\kern-.125emX}}

%% file: sections/abstract.tex
\begin{abstract}
In this paper we provide an overview of the programmed instructions approach for the purpose of quantum software education. The article presents the programmed instructions method and recent successes in STEM fields before describing its operating mode. Elements tackled include the core components of programmed instructions, its behavioural roots and early use as well as adaptation to complex STEM material. In addition, we offer recommendations for its use in the specific context of quantum software education and provide one example of PI-based instruction for the notion of entanglement. The aim of this work is to provide high-level guidelines for incorporating programmed instructions in quantum education with the goal of disseminating quantum skills and notions more efficiently to a wider audience.
\end{abstract}

%% file: sections/intro.tex
\section*{Introduction}

Quantum technologies (QT) is considered, by a growing number of public and private bodies alike\cite{fox2020preparing, kaur2022defining}, to be the next technological revolution. As such, over the course of the last thirty years, the once secluded academic field of research has seeped into industry and made its way to undergraduate -and even in some cases high-school- courses, increasing the momentum of the field. This fast-growing spread bears testimony to the crucial role QTs have been increasingly acknowledged to play. In this context, the magnitude of the workforce required to fill the different roles in QTs has considerably grown, making the need for quantum training and education a priority as recognized by several governments \cite{kaur2022defining}.  On top of STEM-related concepts of programming, linear algebra, and probabilities, come phenomena best described in the language of quantum mechanics such as interference, entanglement, tunnelling and superposition. The variety of concepts invites a vast range of backgrounds ranging from physics to mathematics by way of computer science and software engineering.

Educators typically teach these quantum mechanical phenomena, crucial to quantum computations, in theory-heavy, math-heavy ways with traditional methods, relying on lectures (live or recorded) and some form of slides/blackboard vastly based on didactic teaching. An alternative and more promising approach to this is behavioural education approaches such as Programmed Instructions (PI) \cite{fernald1991programmed}, personalised system of instruction \cite{paiva2017intelligent} , direct instruction \cite{slocum2021features}, computer-assisted instruction  \cite{rigney2020headsprout, root2021towards}, as well as more clinically oriented approaches such as applied behaviour analysis \cite{cooper2007applied}. The essence of behavioural education is based on Skinner's reinforcement principles \cite{twyman2019programmed} and its application has been proven effective in a broad range of populations such as children \cite{boden2000programmed}, college students \cite{sadykov2023systematic}, and people with neurodiverse populations \cite{peters2011meta}, with contents such as early reading and writing \cite{rigney2020headsprout}, academic contents such as chemistry \cite{kurbanoglu2006programmed}, statistics \cite{harrington1999teaching}, physiology \cite{stanisavljevic2013application}, and medicine \cite{Hui2009-kr} as well as computer programming \cite{emurian2000learning}.  On top of being well-adapted to promote learner involvement and success, these approaches are also highly compatible with computerised programs \cite{root2021towards} \cite{harrington1999teaching}.

Because the \textit{quantumrush} is still in its infancy, now is a favourable time for fundamental decisions to be taken regarding the way that knowledge is spread and effectively taught. The primary objective of this paper is to present an overview of the PI teaching/learning paradigm with a particular emphasis on its use in STEM fields and propose its use in educational programs for quantum technologies. By doing so, the authors hope to strengthen the discussion within the quantum community about how content is delivered to students and encourage the adoption of this evidence-based educational approach. We describe the core concepts of PI and provide an easy-to-use checklist for any course content creator to follow to ensure compliance with the principles of PI. In addition, we provide a detailed example of how to use PI in the context of teaching the concept of entanglement from a quantum software development perspective. The article is organised as follows: in section 1 we introduce programmed instructions. In section 2 we give a brief overview of current \tnl approaches in quantum education (QEd). In section 3 we bring together PI and QEd and finally in section 4 we propose an example use of PI for QEd with the notion of entanglement\cite{aiello2021achieving}.

%% file: sections/what_is_pi.tex
\section{What are programmed instructions?}

PI has its origin in psychology, more specifically in the behavioural approach.  The behavioural approach (or behaviourism) in psychology is characterised by its focus on three key aspects. The first is the measurement of observable phenomena -the dependent variable-, usually the behaviour of an organism as opposed to the measurement of concepts such as cognition and personality. Second,  manipulation of external events or environmental stimuli as the independent variables. That is, behaviourists identify causes of behaviour change in the organism's environment as opposed to the organism's internal state such as motivation and thoughts. Lastly, behavioural psychology seeks a causal relationship between behaviour and environmental stimuli. Unlike other disciplines in psychology, the behavioural approach is thus characteristically pragmatic and its scientific findings are readily applicable to applied settings.

The concept of \textit{operant conditioning}, coined by Skinner, is key to behavioural educational approaches \cite{skinner1968technology}. This concept focuses on the modification of behaviours of organisms through environmental feedback. In other words, operant conditioning looks at how organisms learn based on the positive and negative feedback (reinforcement) they receive from their environment. Skinner discovered that the best way to teach a learner a complex skill was to set up small, achievable goals and gradually shape up the learner's behaviour using reinforcement. 
In 1968, Skinner proposed\cite{skinner1968technology} a \textit{teaching machine} as an alternative to didactic teaching. Skinner pointed out that in didactic teaching, a teacher is a presenter of instruction, who gives a group of students a set of instructions at once. Students are required to learn a set of information on their own. On the other hand, in his idea of the teaching machine, a teacher is a programmer of instructions who \textit{programs} the set of instructions for each individual so that every student can follow and progress through instructions one by one at their own pace. This teaching machine had been extensively researched from the '60s to the '80s and showed early success \cite{benjamin1988history}. It is now known as programmed instructions or PI.   

Despite its early success, PI was not integrated into mainstream educational programs. Main criticisms toward PI didn't focus on its effectiveness as PI remained typically more effective than traditional didactic teaching\cite{benjamin1988history}. The focus of criticism was on its \textit{social validity}. Educators and philosophers were at that time concerned with PI's automated teaching and viewed it as lacking love and social interaction between a teacher and a student and making students isolated \cite{root2021towards}. In recent years, the public's attitude toward PI has changed as personal computers became more readily available and online educational materials became common \cite{root2021towards}.

%% file: sections/pi_for_qed.tex
\section{Programmed instructions for QEd}

\subsection{A glimpse into the quantum education landscape}\label{sec:qed}

The last century has seen the number of teaching approaches soar in education. From the teacher-centered didactic instruction to Montessori \cite{dohrmann2007high} by way of inquiry-based and constructivism, knowledge flow has received considerable attention. However, the further away one goes from primary school to higher education, the less likely it is that the teaching method differs from the widespread \textit{lecture-style didactic teaching}. 

As detailed in the excellent review by Kaur \textit{et.al.}\cite{kaur2022defining} the current quantum education landscape offers a wide variety of resources and approaches. Industry and academia \cite{aiello2021achieving} have put considerable effort into content creation, striving to make quantum education available to a growing audience of non-specialists and early-stage learners \cite{temporao2022teaching}. However, to the best of our knowledge, no quantum educational resources rely clearly on PI so far. Based on the strong evidence of success of PI approaches in STEM fields \cite{kurbanoglu2006programmed, harrington1999teaching, stanisavljevic2013application, Hui2009-kr, emurian2000learning}, its exploitation in QEd appears to be a promising avenue to explore.

\subsection{From didactics to PI}
Shifting from the common didactic approach to a PI approach requires re-framing the knowledge flow and, more than anything, the purpose of its education. 

Established in the late 1980s by Anderson \textit{et.al.} \cite{anderson1991research}, \textit{instructional dimensions} were initially used to differentiate inquiry-based instruction from non-inquiry-based instruction. Instructional dimensions represent a valuable tool for evaluating PI versus non-PI approaches by providing a measuring tool for the aspects which present a shift when switching from one approach to the other. Here we suggest a brief comparative table contrasting PI and non-PI approaches against Anderson \textit{et.al.}'s instructional dimensions.
\input{tables_images/instructional_dimensions}
\subsection{PI components} \label{sec:picomponents}

PI consists of four elements: Behavioural objectives, response requirements, feedback, and mastery criterion.

Behavioural objectives are learning objectives that are operational and objective. The underlying assumption of PI is that the steps involved in learning a complex task, such as quantum computation, can be broken down into a set of tiny, observable incremental steps called frames, that can be specified with sufficient precision \cite{EMURIAN2000395}. Only one frame is presented to a learner at a time and only one piece of information is presented per frame. For example, a typical didactic material would present several pieces of information in one presentation. In a number of standard introductions on qubits, a vast number of concurrent pieces of information are presented at once. Typically a single chapter contains the concept of  information being processed as a series of 0s and 1s, the fact that quantum bits are called qubits, the fact that those obey the rules of quantum mechanics, the idea that qubits will process information in a different way, and so on. On the other hand, in PI, each of these facts would correspond to a single frame and would be presented separately in a sequential manner to a learner. Thus, only one frame would be presented at a time, containing one single new piece of information, and the learner would then be asked to interact with it. 

Response requirements of PI are behavioural and operational. In the context of software engineering, \textit{learning} computer programming can be specified as several types of verbal behaviours such as recognition of items, identification of associations among items, identification of orders among items, as well as production of these items in different written relations \cite{EMURIAN2000395}. The size and amount of verbal response requirements will increase as a learner masters each step. For example, after a frame containing the piece of information that \textit{a bit is the smallest unit of information there is}, is presented, PI requires a learner to engage in multiple choice questions, fill-in blank statements, as well as to write what a bit is. In typical didactic teaching, a learner is not only provided with many pieces of information at once, but is also asked to engage in very few responses during the lecture, interaction usually coming in much later via various forms of assignments and exams. On the other hand, PI is intensely interactive and forces the learner to engage with pieces of information to facilitate a learner's information acquisition on time. 

Because response requirements are objective and on time, direct and immediate feedback is also prepared for each frame of behavioural objectives. Any type of a learner's responses, be it receptive or expressive, can be evaluated as correct or incorrect immediately. Correctional feedback is also presented alongside with it. Unlike typical didactic teaching where not only assessment of information acquisition but also feedback is delayed, IP provides immediate and specific feedback on the spot to increase a learner's mastery of each piece of information.

The above components of PI allow for individualised, mastery-based progression in learning for every student. Moreover, unlike typical didactic teaching, PI is self-paced and tailored to a learner's own learning style. A learner can initiate the program whenever and wherever suitable for them and the information is given frame by frame according to the learner's mastery level. PI also requires a learner to emit various responses, which facilitates active learning rather than passive learning. In addition, information is not presented to a group of people at once as would be with a live lecture but on an on-demand basis. PI materials are always individually adjusted to a learner's level of proficiency in a given area and the progress is both gradual and individual. 

Lastly, each behavioural objective, response requirement, and mastery criterion within a given PI set is programmed based on the specific overall goal of each PI set. This means that even if the topic of study is the same, different pursuits would lead to different ways of building the PI material. For instance, the concept of quantum entanglement would be presented in different ways depending on the audience it is targeting. Where an experimental physicist might need to acquire mathematical equations and experimental devices on entanglement, a theoretical computer scientist needs the algorithmic possibilities that can be unlocked through entanglement. For a software engineer, PI set would focus on the best way to code entanglement in systems. In the case of teaching quantum software engineers, each behavioural objective, frame, and mastery criterion would be geared toward the overall goal of a learner mastering writing adequate code.

\subsection{PI checklist} \label{sec:piChecklist}
To facilitate adoption by educators, we propose a brief checklist of the core elements of PI. When building educational sequences or designing course material, this simple checklist can be used to evaluate how in-line one's program is with PI. This is by no way exhaustive and we encourage readers to explore further\footnote{references given in sec \ref{sec:qed} for instance} to build a deeper view of each element.

\begin{itemize}
    \item Is there a set of finite, specific, clearly designed behavioural objectives -distinct pieces of information separated in frames- for the learners?
    \item Are there detailed response requirements for each behavioural objective -a set of specific learning objectives as measurable by concrete learner actions-?
    \item Is direct, objective, and immediate feedback \cite{jaehnig2007feedback} available to each frame? I.e. is there a communication channel that can provide feedback to the student based on their specific responses programmed in the PI? A mechanism to comment on the learner's response, providing some degree of personalised guidance as opposed to group-based remarks?
    \item Is the progression offered to each learner individualised?
    \item Is the progression offered to each learner based on mastery of the subject \cite{vargas1991programmed}? I.e. is mastery of the subject evaluated gradually with potential remediation offered throughout?
\end{itemize}

%% file: tables_images/instructional_dimensions.tex
\begin{table*}[t]
\centering
\label{tab:horizon_impact}
\begin{tabular}{l p{7cm} p{7cm}}
\toprule
  & PI & non-PI   \\ 
\cmidrule(l){2-3}
subject matter & unchanged & unchanged \\
activity demands & behaviour-oriented, personalised & konwledge-oriented, shared by group  \\
instructional format & specific, behaviour-oriented & more general, knowledge-oriented \\
grouping & small conceptual units & potentially large chapters including numerous sub-topics \\
time management & feedback obtained immediately & feedback delayed, possibly for days \\
teacher-learner interaction & direct interaction of learner with knowledge at time of learning & teacher acts as a compulsory interface between learner and knowledge at time of learning \\
\bottomrule
\end{tabular}
\caption{PI versus non-PI approaches based on instructional dimensions}
\end{table*}

%% file: sections/example_pi.tex

\section{Example use-case : a programmed instructions approach to entanglement for early-stage quantum software engineers}
In this section, we provide a toy-sized example of a PI sequence designed to teach early-stage generalist engineers how to code entangled circuits using Qiskit\footnote{Note that we rely on Qiskit as an example quantum programming language but encourage readers to explore the vast variety of existing quantum programming languages.}. As PI is a behavioural approach that therefore relies heavily on quantifiable, observable responses as results, defining a clear objective and detailing a precise target audience is paramount. The over-arching goal of this mini-sequence is therefore defined as teaching generalist engineers with \textit{some} previous exposure to quantum programming which we detail below. Note that our approach here relies purely on the engineering aspects of quantum programming and makes no assumption about theoretical proficiency which, from the PI perspective, is considered as a separate piece of knowledge irrelevant in this case. While a theoretical computer scientist or physicist would require mathematical equations, a software engineer might not need them as readily. Throughout our example, we use the checklist provided in sec.\ref{sec:piChecklist} to illustrate its purpose.

\subsection{Prerequisites}\label{sec:prerequisites}
As PI focuses on single conceptual units and their gradual assimilation, some other important and necessary notions are here treated as prerequisites assuming that a learner acquired them in prior sessions with corresponding PI materials. We assume familiarity with the following concepts:
\begin{itemize}
    \item \textbf{Quantum circuits} : -using the Qiskit representation- how to read and interpret them, how to instantiate one.
    \item \textbf{Single qubit gates} : what they are, how they are used, how to code the basic set of gates X, Z and H.
    \item \textbf{Two qubit gates} : what they are, how they are used, how they act on qubits as well as how to code a set of basic gates such as CX and CZ.
    \item \textbf{Superposition} : the basic concept and implications, how to generate superposition with an H gate for instance.
    \item \textbf{Bracket notation} : how to read kets, bras and multiple qubit systems.
    \item \textbf{Conventional initial state} : fact that conventionally, if no further specifications are communicated, a circuit is initialised in the state $\ket{00\ldots0}$
\end{itemize}

\subsection{PI checklist}
\subsubsection{Behavioural objectives}
Each frame has a specific behavioural objective (one piece of information to teach).

\subsubsection{Response requirements}
After being exposed to the PI set about entanglement, the learner possesses the ability to:
\begin{itemize}
    \item Discriminate between gate families that allow or don't allow the creation of entanglement (single vs multiple-qubit gates).
    \item When required to entangle two qubits, know how to choose a gate to do so.
    \item Identify whether a given quantum system is entangleable or not based on the number of qubits it has.
    \item Given a diagrammatic or code representation of a circuit, differentiate between circuits displaying entanglement and circuits not displaying entanglement.
    \item Given a diagrammatic or code representation of an entangled circuit, list the different entanglement relations between qubits, tell the difference between those which are entangled and those which are not, perceive which qubits are entangled with which others.
    \item Given a diagrammatic representation, write the code corresponding to said circuit.
    \item Given a code representation, draw the corresponding diagram.
    \item Given a set of requirements and constraints, write a piece of code and/or draw a circuit diagram which fulfils those requirements and respect those constraints.
\end{itemize}

\subsubsection{Feedback structure and timing}
Assuming these frames were presented on a learning application or software, feedback would be given immediately, after each frame. Any incorrect answer would trigger a visualisation of the correct answer and the learner would be prompted again with the frame they had just got wrong first, before circling back to the other frames.

\subsubsection{Individualised progression}
Assuming, again, that this content would be presented through a learning application of software, the progression would be individualised for each learner by construction.

See below for example frames.
\input{tables_images/frames18}
\input{tables_images/frames916}
\input{tables_images/frames1724}
\input{tables_images/frames25}

%% file: tables_images/frames18.tex
\begin{figure*}
\centering

\captionbox{Frame 1}
[.475\textwidth]{\frame{\includegraphics[width=.45\textwidth]{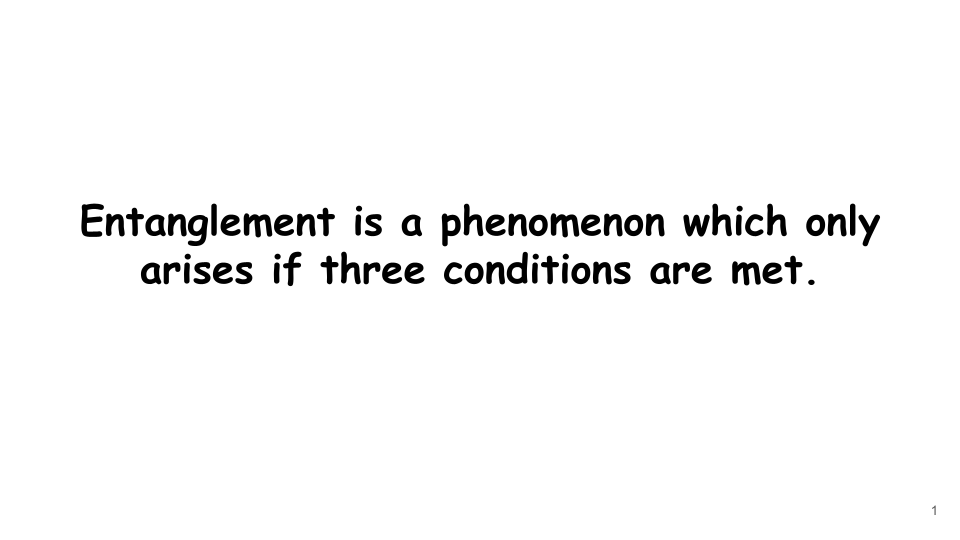}}}
\captionbox{Frame 2}
[.475\textwidth]{\frame{\includegraphics[width=.45\textwidth]{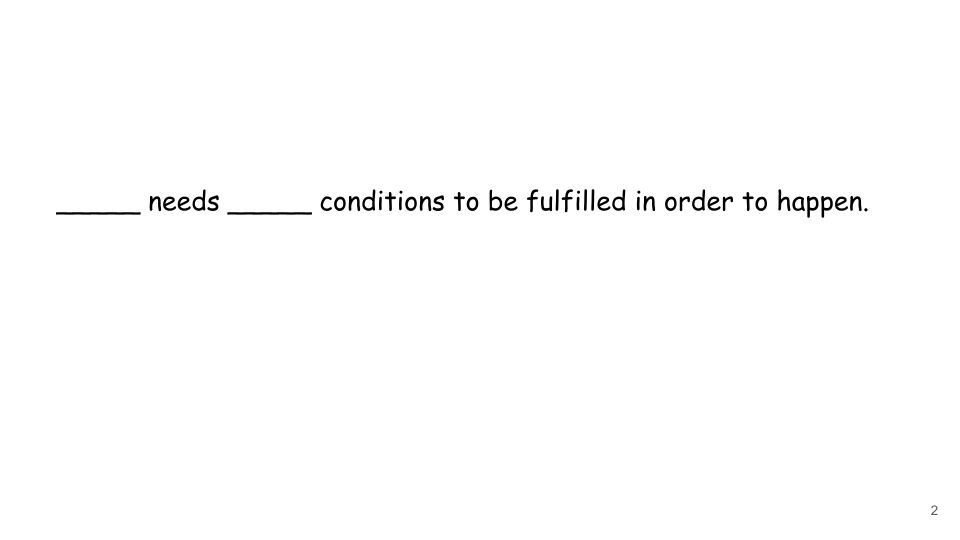}}}

\captionbox{Frame 3}
[.475\textwidth]{\frame{\includegraphics[width=.45\textwidth]{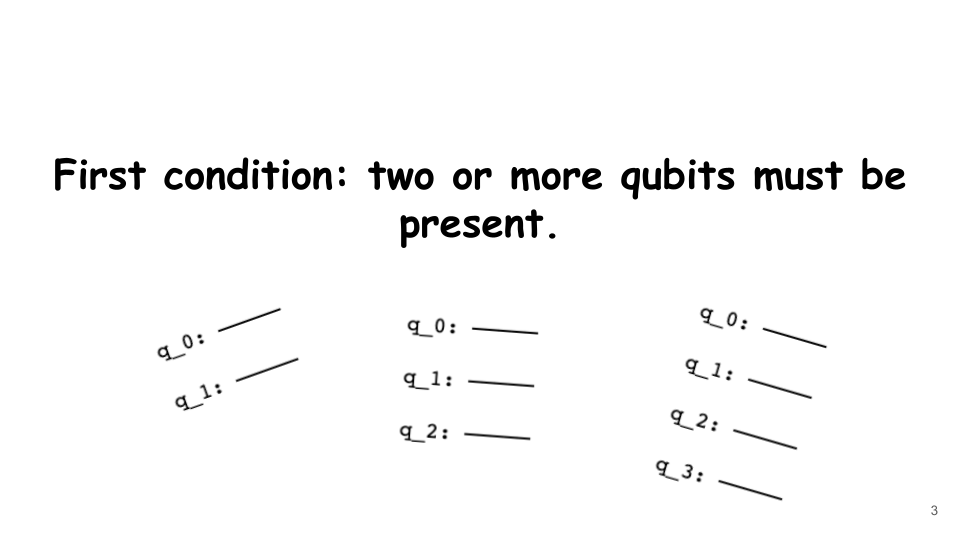}}}
\captionbox{Frame 4}
[.475\textwidth]{\frame{\includegraphics[width=.45\textwidth]{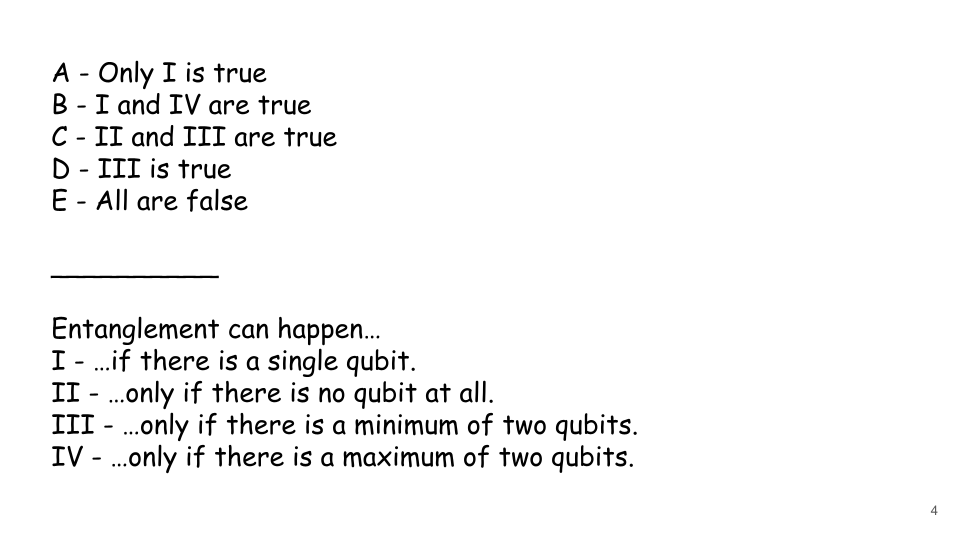}}}

\captionbox{Frame 5}
[.475\textwidth]{\frame{\includegraphics[width=.45\textwidth]{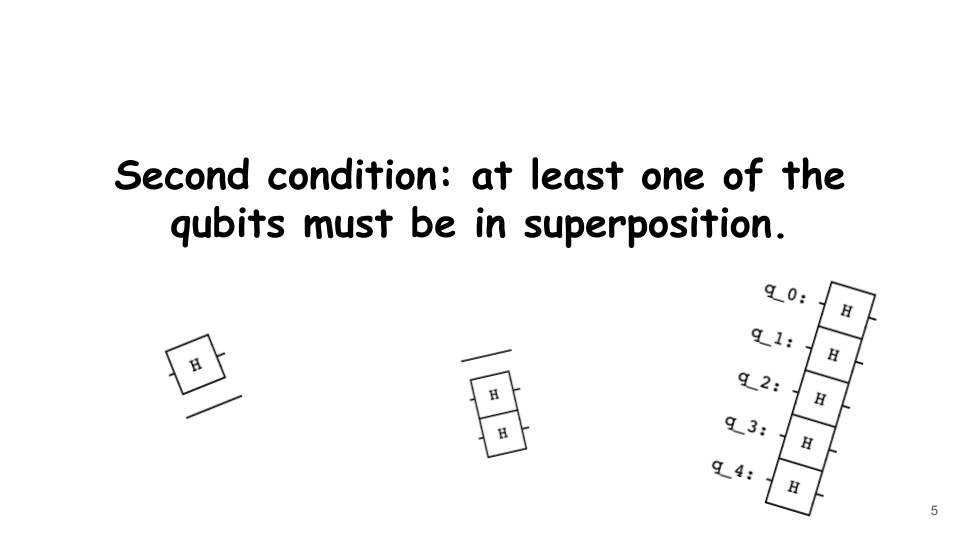}}}
\captionbox{Frame 6}
[.475\textwidth]{\frame{\includegraphics[width=.45\textwidth]{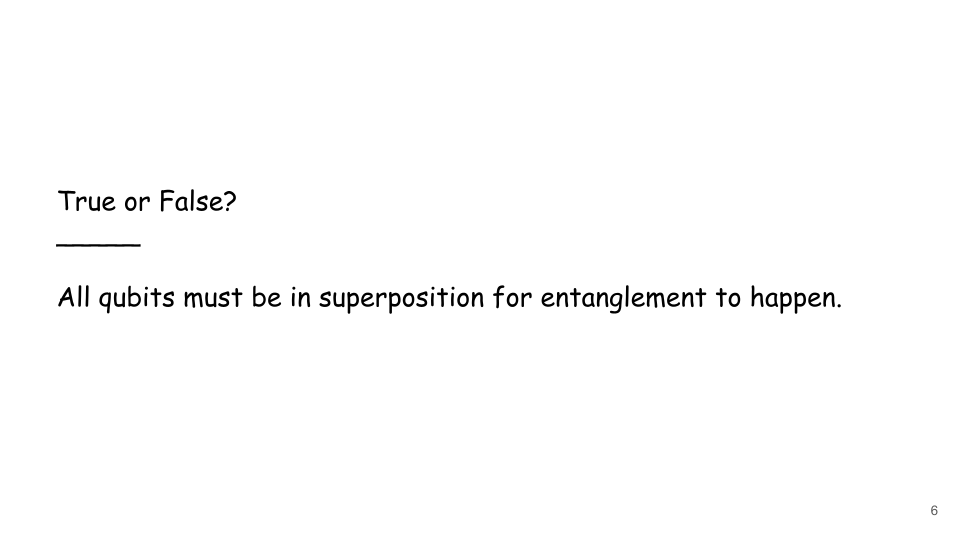}}}

\captionbox{Frame 7}
[.475\textwidth]{\frame{\includegraphics[width=.45\textwidth]{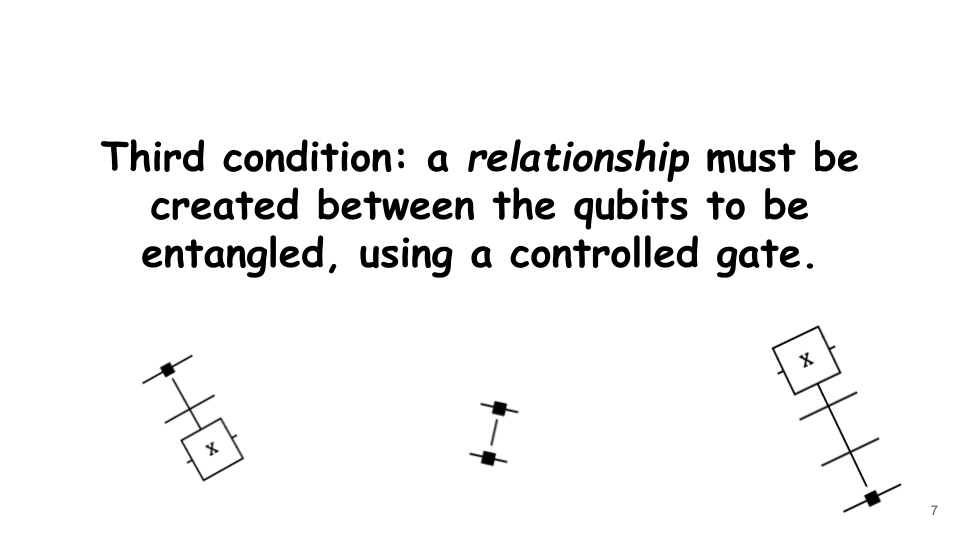}}}
\captionbox{Frame 8}
[.475\textwidth]{\frame{\includegraphics[width=.45\textwidth]{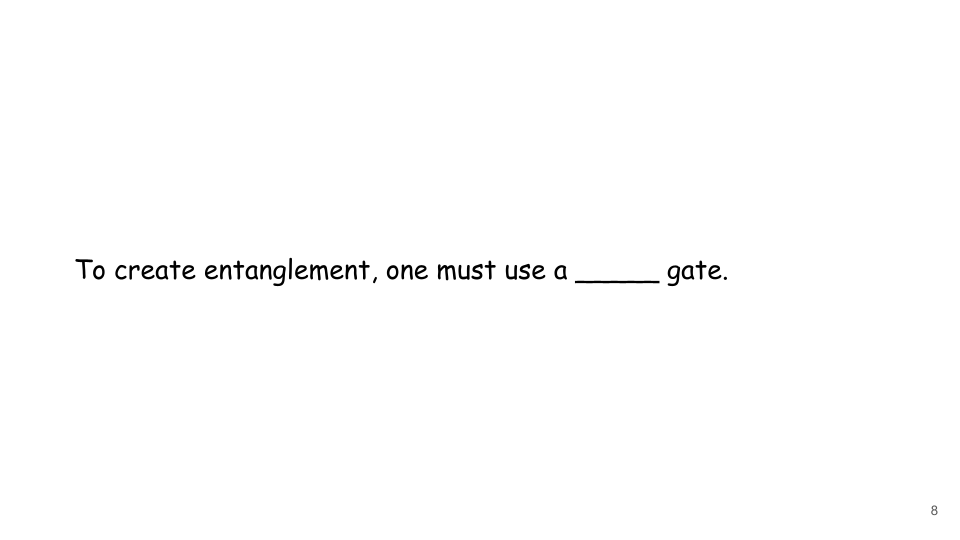}}}

\end{figure*}

%% file: tables_images/frames916.tex
\begin{figure*}
\centering

\captionbox{Frame 9}
[.475\textwidth]{\frame{\includegraphics[width=.45\textwidth]{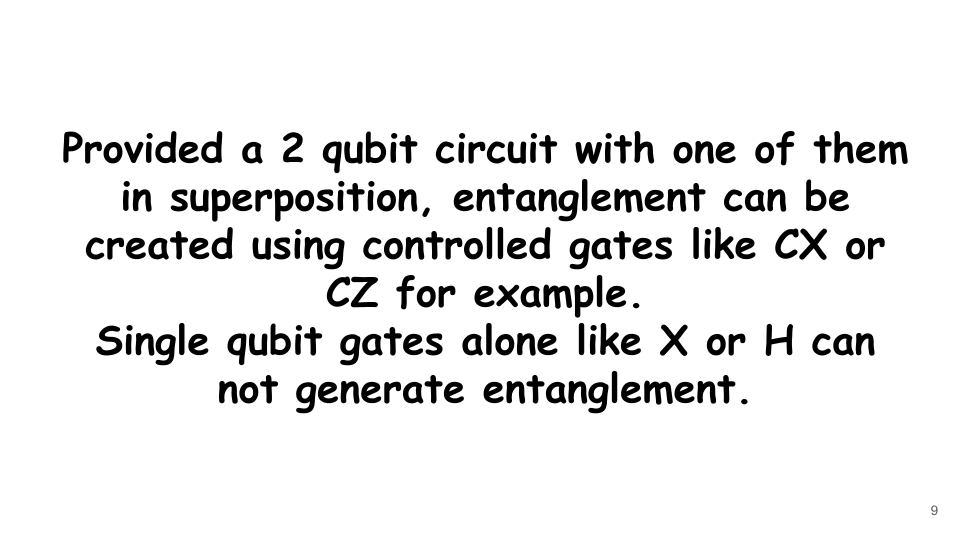}}}
\captionbox{Frame 10}
[.475\textwidth]{\frame{\includegraphics[width=.45\textwidth]{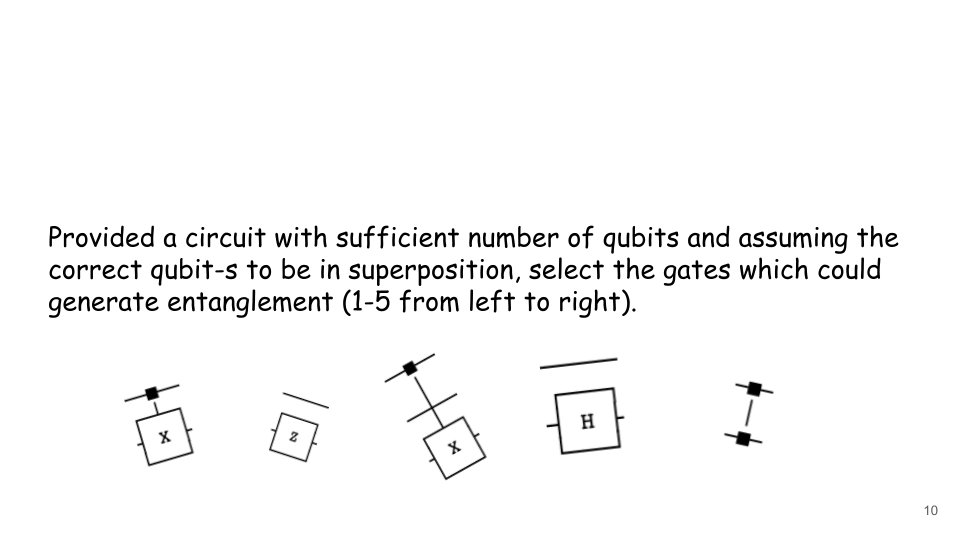}}}

\captionbox{Frame 11}
[.475\textwidth]{\frame{\includegraphics[width=.45\textwidth]{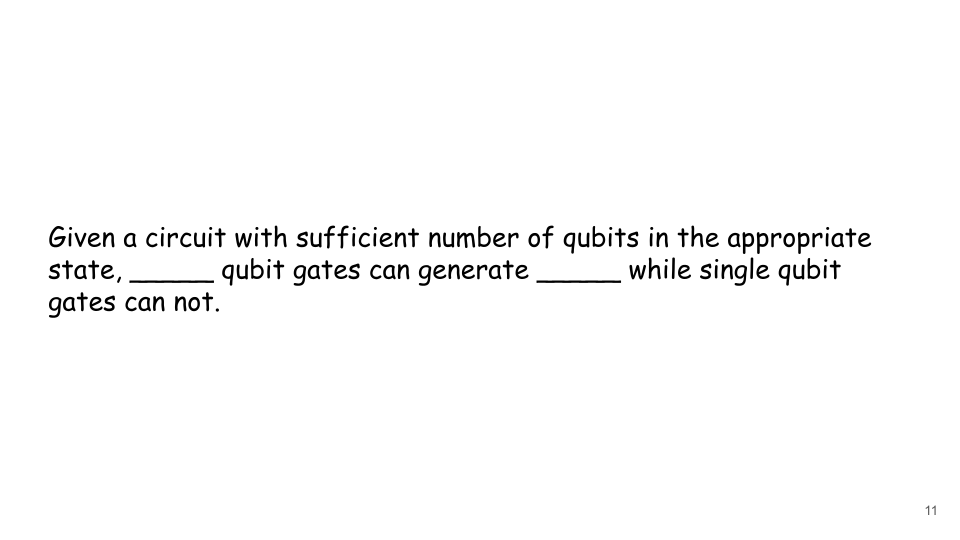}}}
\captionbox{Frame 12}
[.475\textwidth]{\frame{\includegraphics[width=.45\textwidth]{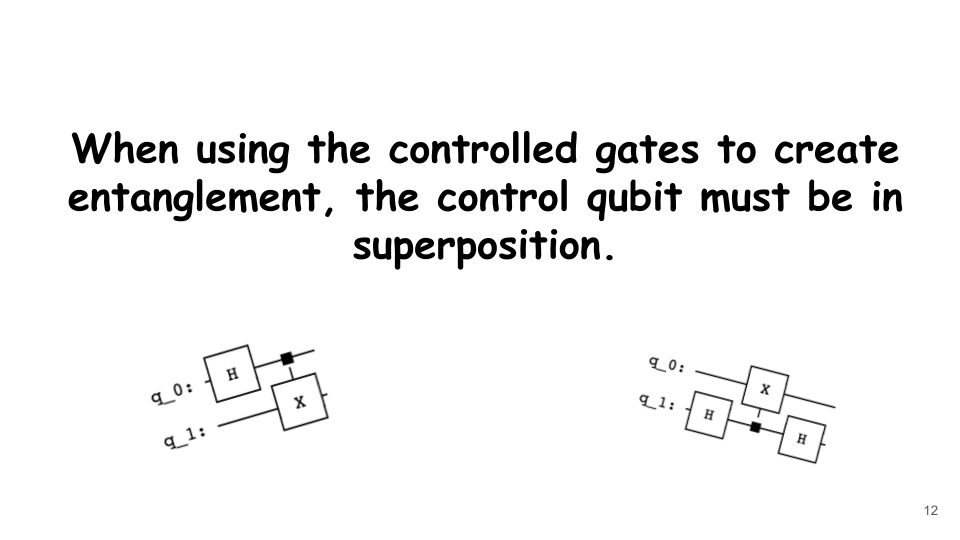}}}

\captionbox{Frame 13}
[.475\textwidth]{\frame{\includegraphics[width=.45\textwidth]{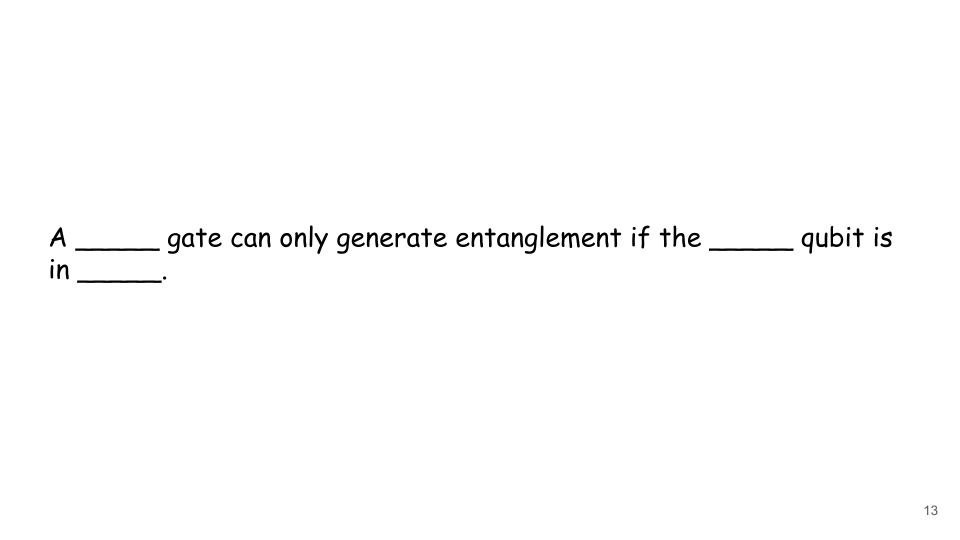}}}
\captionbox{Frame 14}
[.475\textwidth]{\frame{\includegraphics[width=.45\textwidth]{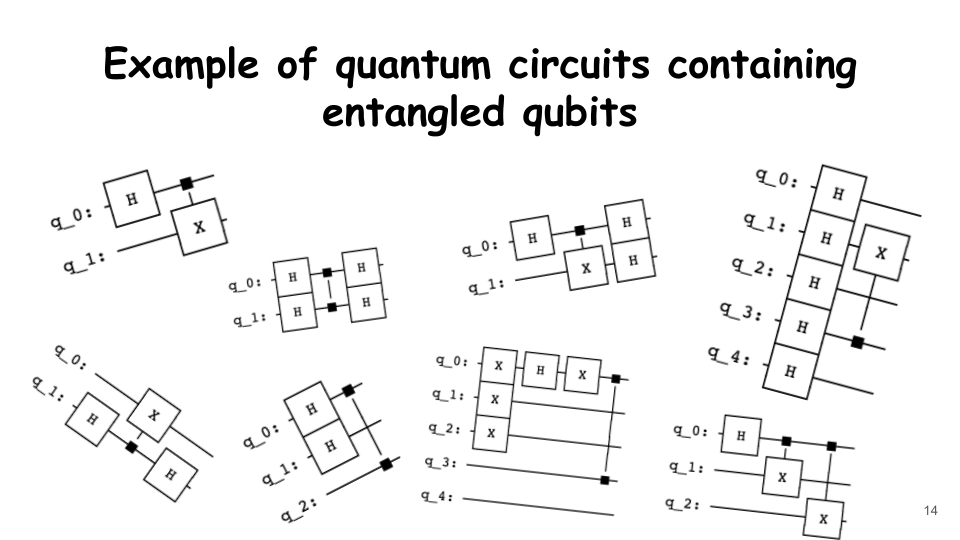}}}

\captionbox{Frame 15}
[.475\textwidth]{\frame{\includegraphics[width=.45\textwidth]{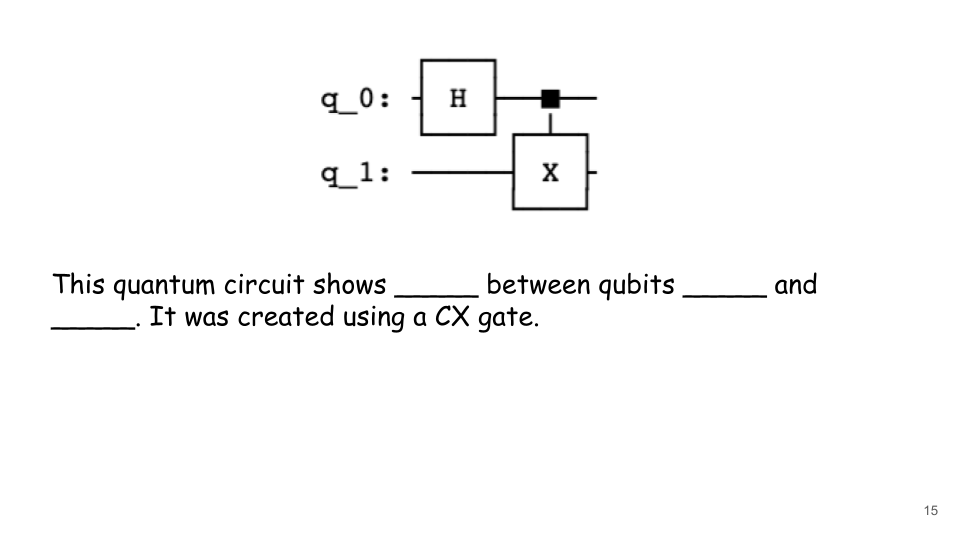}}}
\captionbox{Frame 16}
[.475\textwidth]{\frame{\includegraphics[width=.45\textwidth]{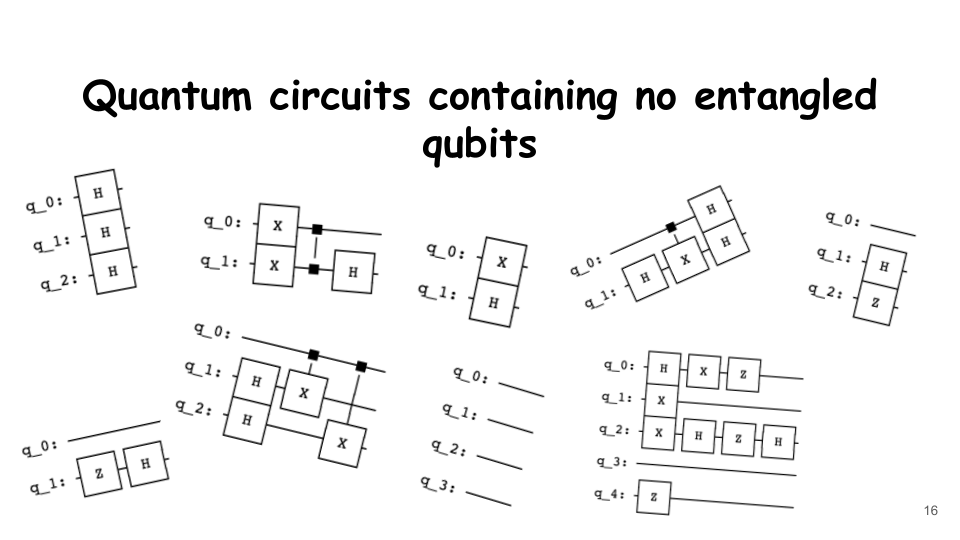}}}

\end{figure*}

%% file: tables_images/frames1724.tex
\begin{figure*}
\centering

\captionbox{Frame 17}
[.475\textwidth]{\frame{\includegraphics[width=.45\textwidth]{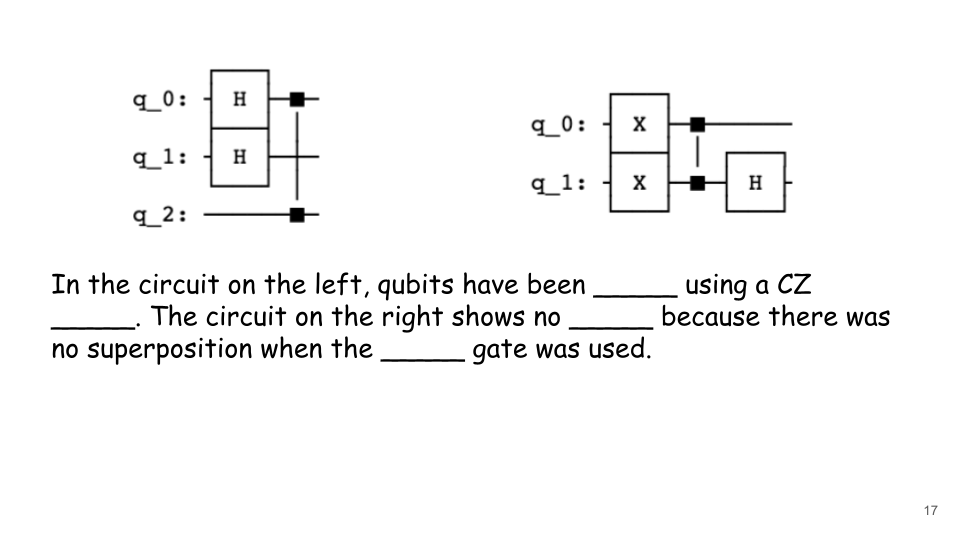}}}
\captionbox{Frame 18}
[.475\textwidth]{\frame{\includegraphics[width=.45\textwidth]{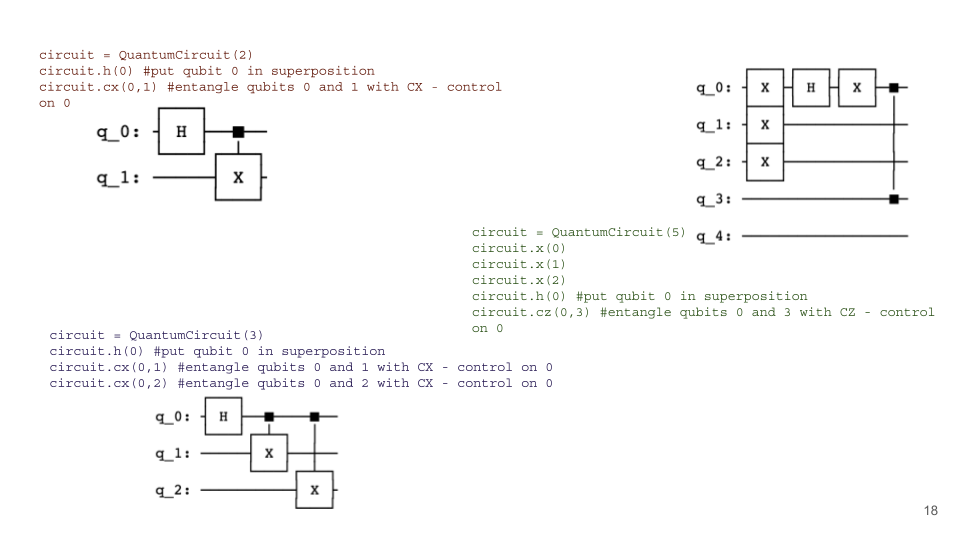}}}

\captionbox{Frame 19}
[.475\textwidth]{\frame{\includegraphics[width=.45\textwidth]{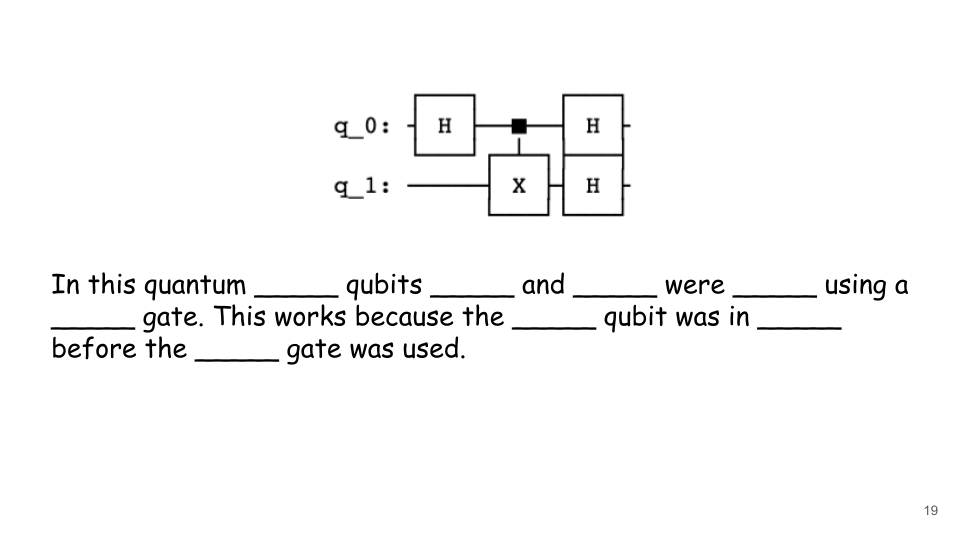}}}
\captionbox{Frame 20}
[.475\textwidth]{\frame{\includegraphics[width=.45\textwidth]{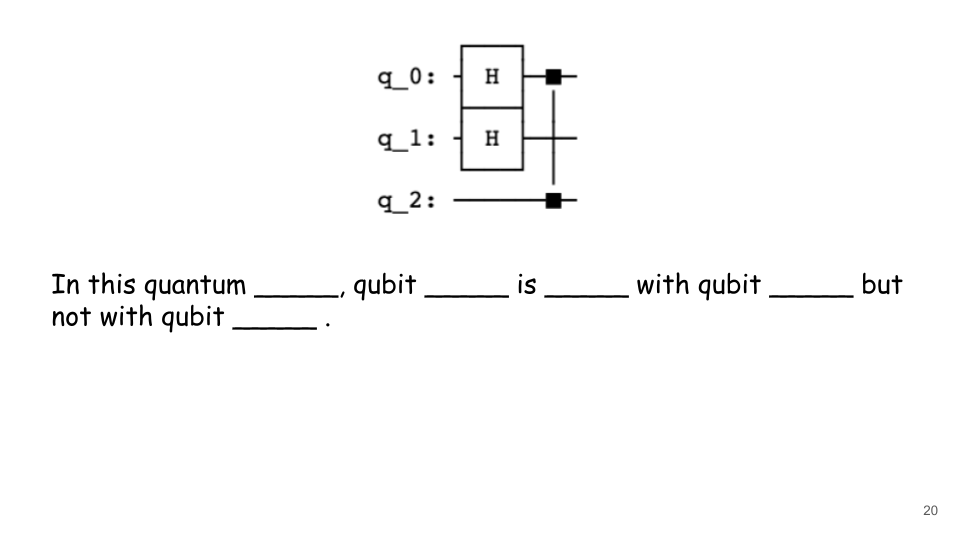}}}

\captionbox{Frame 21}
[.475\textwidth]{\frame{\includegraphics[width=.45\textwidth]{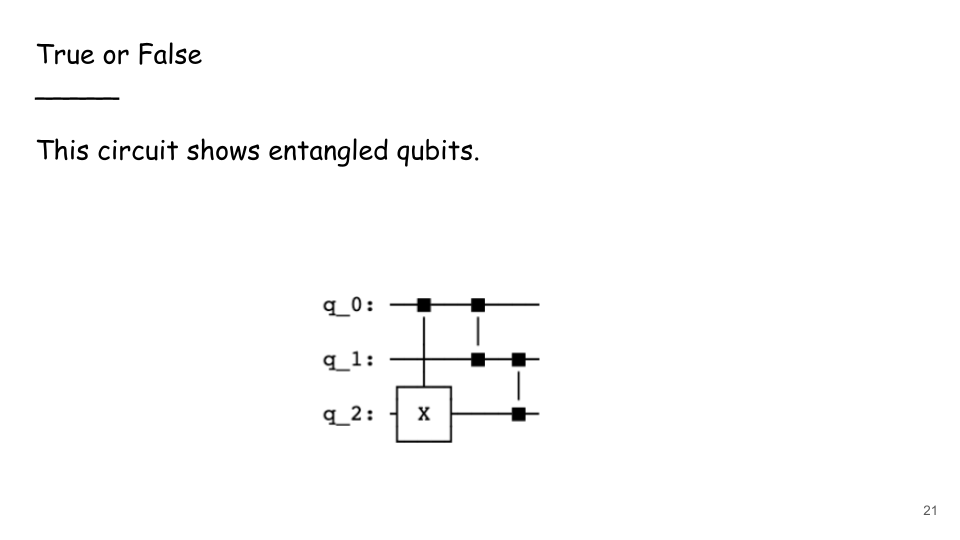}}}
\captionbox{Frame 22}
[.475\textwidth]{\frame{\includegraphics[width=.45\textwidth]{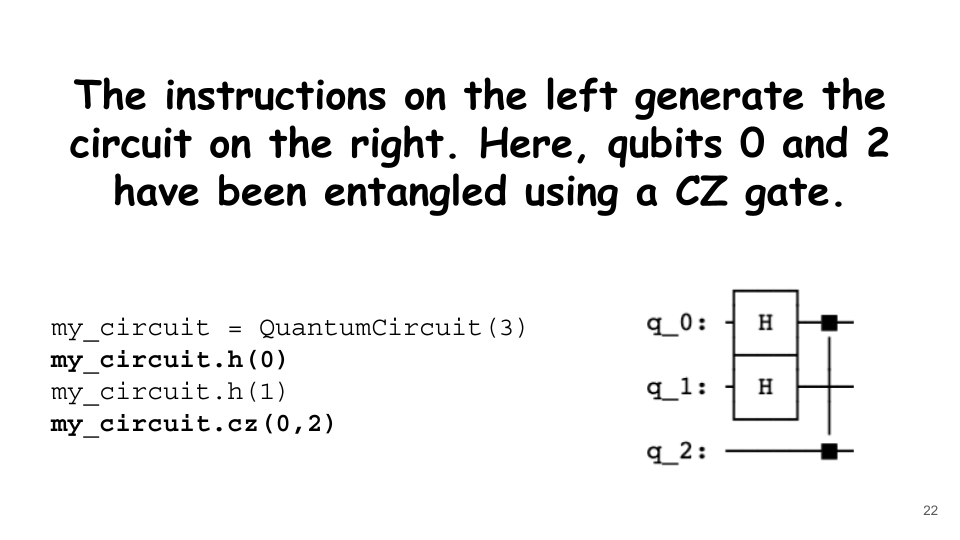}}}

\captionbox{Frame 23}
[.475\textwidth]{\frame{\includegraphics[width=.45\textwidth]{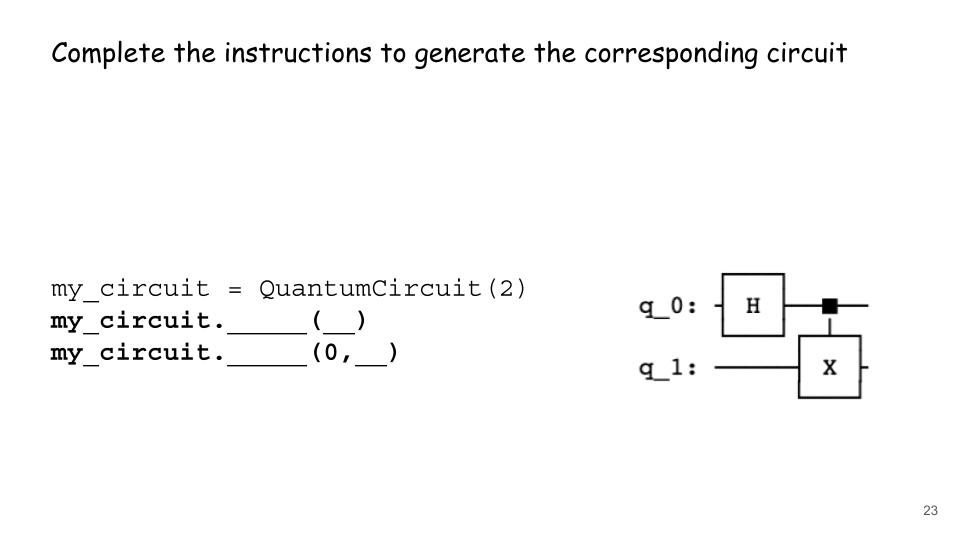}}}
\captionbox{Frame 24}
[.475\textwidth]{\frame{\includegraphics[width=.45\textwidth]{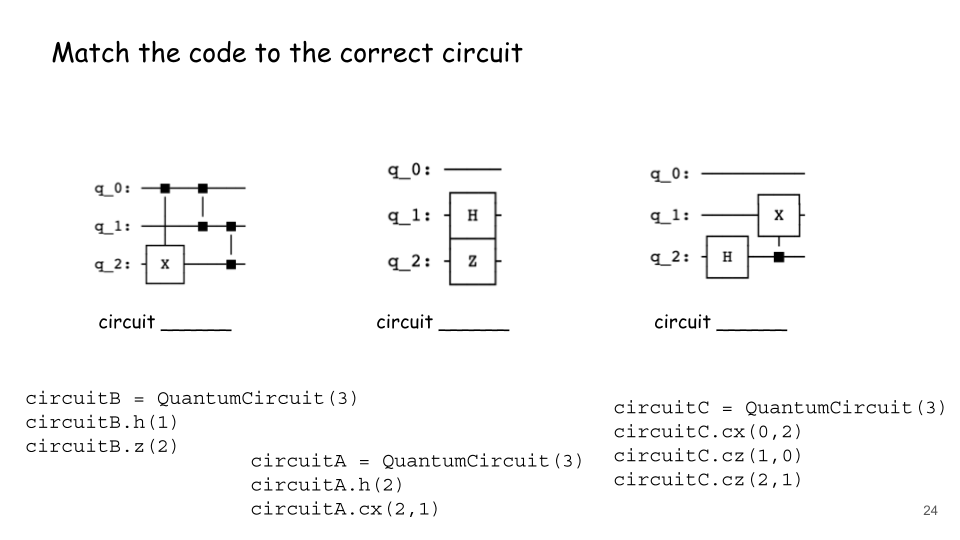}}}

\end{figure*}

%% file: tables_images/frames25.tex
\begin{figure*}
\centering

\captionbox{Frame 25}
[.475\textwidth]{\frame{\includegraphics[width=.45\textwidth]{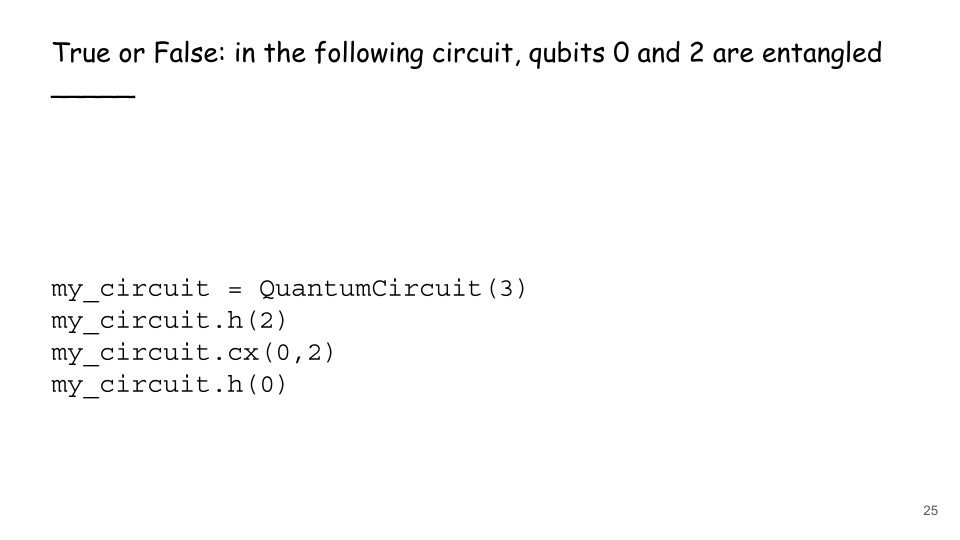}}}
\captionbox{Frame 26}
[.475\textwidth]{\frame{\includegraphics[width=.45\textwidth]{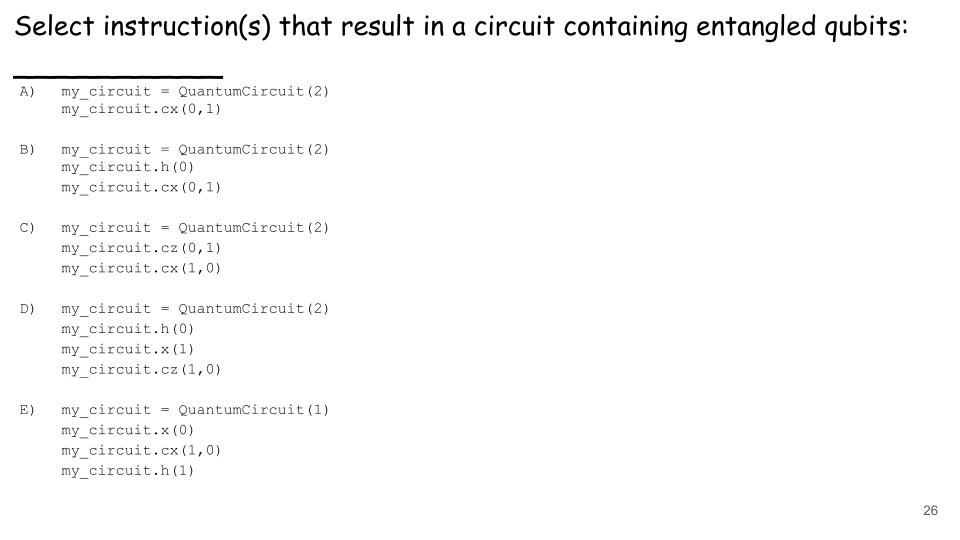}}}

\captionbox{Frame 27}
[.475\textwidth]{\frame{\includegraphics[width=.45\textwidth]{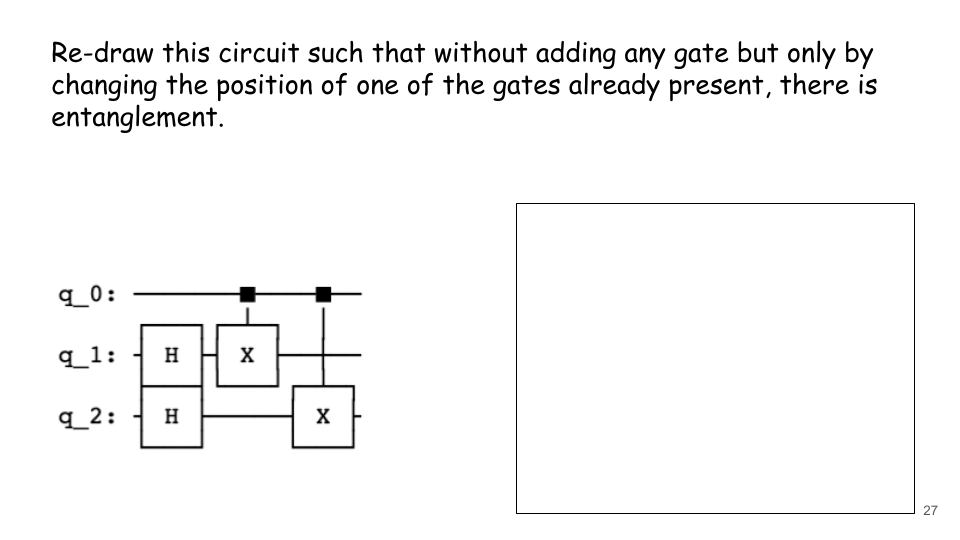}}}
\captionbox{Frame 28}
[.475\textwidth]{\frame{\includegraphics[width=.45\textwidth]{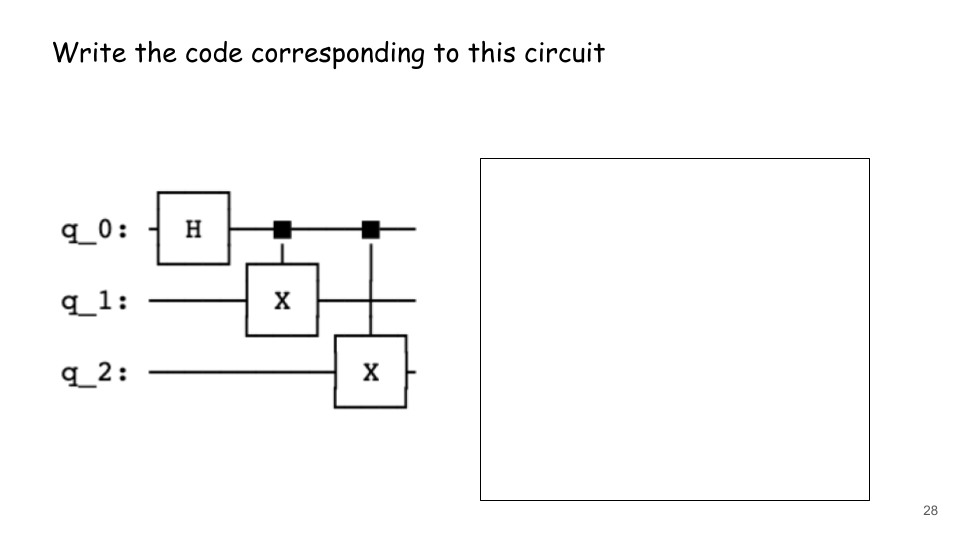}}}

\captionbox{Frame 29}
[.475\textwidth]{\frame{\includegraphics[width=.45\textwidth]{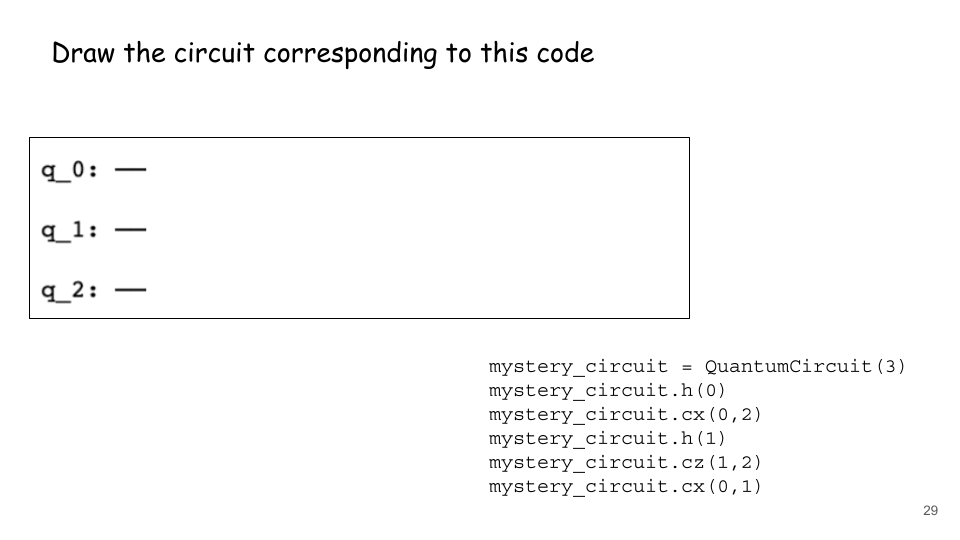}}}
\captionbox{Frame 30}
[.475\textwidth]{\frame{\includegraphics[width=.45\textwidth]{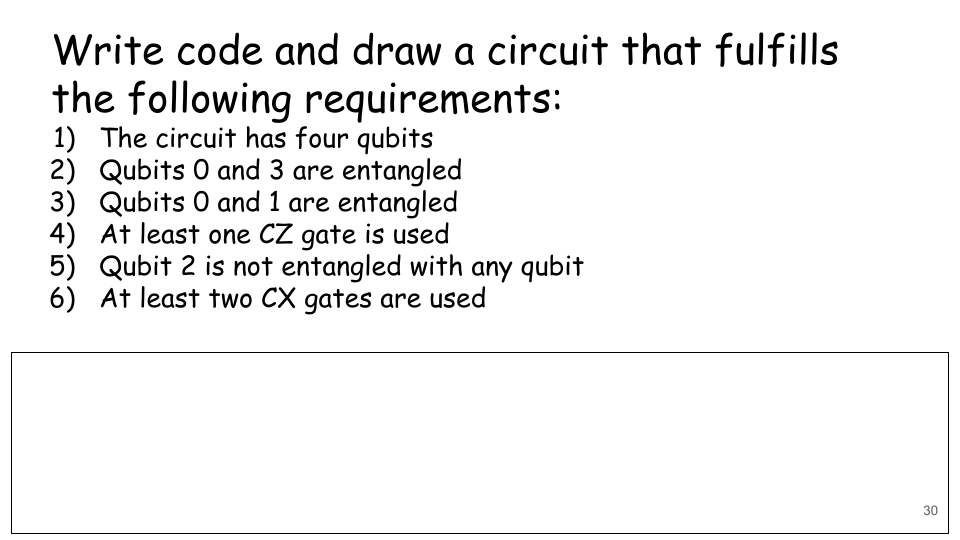}}}

\end{figure*}

%% file: sections/conclusion.tex
\section*{Conclusion}
In this short paper, we gave a brief overview of the programmed instructions teaching paradigm and proposed its potential benefits for quantum education in general. We also detailed an example use case by introducing the notion of entanglement for quantum software engineers through a set of programmed instructions frames. We hope it will prove useful to educators and course-designers in their efforts to create ever more efficient approaches to knowledge transfer.

%% file: sections/acknowledgments.tex
\section*{Acknowledgement}
The authors are deeply grateful to Aoi Hayashi from the Okinawa Institute of Science and Technology for his thorough proofreading and insightful suggestions. Authors also thank Filippo Miatto for valuable discussions and the dedicated proofreaders and beta-testers for their remarks which proved invaluable in bettering this work. Finally, authors thank the anonymous reviewers for their detailed and helpful feedback.

%% file: sections/annex.tex
\section*{Annex}

\subsection*{Solutions to frames}
\begin{itemize}
    \item \textbf{Frame 2} : entanglement, 3
    \item \textbf{Frame4 4} : D
    \item \textbf{Frame 6} : False
    \item \textbf{Frame 8} : controlled / two-qubit
    \item \textbf{Frame 10} : 1,3,5
    \item \textbf{Frame 11} : two, entanglement
    \item \textbf{Frame 13} : controlled / two-qubit, control, superposition
    \item \textbf{Frame 15} : entanglement, 0, 1
    \item \textbf{Frame 17} : entangled, gate, entanglement, controlled / two-qubit
    \item  \textbf{Frame 19} : circuit, 0, 1, entangled, CX / CNOT, control, superposition, controlled / two-qubit
    \item \textbf{Frame 20} : circuit, 0, entangled, 2, 1
    \item \textbf{Frame 21} : False
    \item \textbf{Frame 23} : h(0), cx(0,1)
    \item \textbf{Frame 24} : from left to right, first circuit = C, second circuit = B, third circuit = A
    \item \textbf{Frame 25} : False
    \item \textbf{Frame 26} : B
    \item \textbf{Frame 27} : see fig. \ref{fig:sol27} \begin{figure}[h]
        \centering
        \includegraphics[scale=0.7]{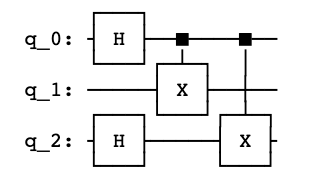}
        \caption{Potential solution to frame 27}
        \label{fig:sol27}
    \end{figure}
    \item \textbf{Frame 28} : \texttt{
        circ = QuantumCircuit(3)
        circ.h(0)
        circ.cx(0,1)
        circ.cx(0,2)
    }
    \item \textbf{Frame 29} : see fig. \ref{fig:sol29} \begin{figure}[h]
        \centering
        \includegraphics[scale=0.7]{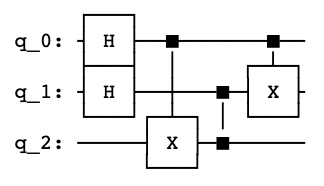}
        \caption{Potential solution to frame 29}
        \label{fig:sol29}
    \end{figure}
    \item \textbf{Frame 30} : this is one \textit{possible} solution out of many \texttt{
        circ = QuantumCircuit(4)
        circ.h(0)
        circ.h(1)
        circ.cx(0,1)
        circ.cz(1,3)
        circ.cx(0,3)
    }
\end{itemize}

\subsection*{Programmed instructions objective formulation guidance}
Aware that re-framing knowledge in a behavioural manner and steering clear of objectives such as \textit{know} or \textit{understand} can be a challenge initially, we provide a non-exhaustive list of descriptors for behavioural objectives\cite{gronlund2004writing}. This list is a mere kick-starter to allow the interested readers to develop their own behavioural objectives for PI and we encourage all interested in exploring the extensive possibilities of natural language and action verbs.
\begin{itemize}
    \item \textbf{Information acquisition} : define, describe, list, detail, report, record, \ldots
    \item \textbf{Information communication} : discuss, explain, express, synthesise, \ldots
    \item \textbf{Information analysis} : analyse, identify, recognise, compare, contrast, \ldots
    \item \textbf{Information leveraging} : score, measure, evaluate, criticise, decide, plan\ldots
\end{itemize}